\documentclass[prl,twocolumn,twoside,showpacs,superscriptaddress]{revtex4}
\usepackage{graphics,amssymb,amsmath,epsf}
\epsfclipon

\newcommand{\be}{\begin{equation}}
\newcommand{\ee}{\end{equation}}

\begin{document}

\title{Solutions of renormalization group flow equations with full momentum dependence}

\author{F. Benitez}
\affiliation{Instituto de F\`{\i}sica, Faculdad de Ingenier\'{\i}a, 
Universidade de la Rep\'ublica, 11000 Montevideo, Uruguay}

\author{J.-P. Blaizot }
\affiliation{IPhT, CEA-Saclay, 91191 Gif-sur-Yvette, France}

\author{H. Chat\'e}
\affiliation{Service de Physique de l'Etat Condens\'e, CEA-Saclay, 91191 Gif-sur-Yvette, France}
\author{B. Delamotte}
\affiliation{LPTMC, CNRS-UMR 7600, Universit\'e Pierre et Marie Curie, 75252 Paris, France}
\author{R. M\'endez-Galain}
\author{N. Wschebor}
\affiliation{Instituto de F\`{\i}sica, Faculdad de Ingenier\'{\i}a, 
Universidade de la Rep\'ublica, 11000 Montevideo, Uruguay}

\date{\today}

\begin{abstract}
We demonstrate the power of a recently-proposed approximation scheme for 
the non-perturbative renormalization group that gives access to correlation 
functions over their full momentum range. 
We solve numerically the  leading-order flow equations obtained within 
this scheme, and compute the two-point functions of the $O(N)$ theories at 
criticality, in two and three dimensions.
Excellent results are obtained  for both universal and non-universal 
quantities at modest numerical cost.
\end{abstract}


\maketitle

The renormalization group, in its non-perturbative version \cite{Wetterich92,Berges00} 
(also referred to as the exact renormalization group), provides  a general formalism giving 
access, for arbitrary coupling strength, to a whole set of physically important quantities,  
universal as well as non-universal \cite{seide99,canet04b}, thermodynamic functions and 
momentum-dependent correlation functions, etc. 
However, most studies within this framework  involve 
approximations that restrict their scope to  the calculation 
of thermodynamical quantities, or correlation functions with vanishing external momenta. 
In order to access the full momentum dependence,
 Blaizot, M\'endez-Galain, and Wschebor (BMW) have recently 
introduced an approximation scheme  which
overcomes this limitation\cite{BMW}. 
In principle, this scheme allows to compute, in all dimensions, at and away from criticality,
both universal and non-universal quantities, as well as momentum-dependent
properties from $p=0$ up to the 
ultra-violet cut-off $\Lambda$ (inverse lattice spacing).

In this Letter, we present the first complete implementation of the 
leading order approximation of the BMW scheme, and demonstrate its power  
by using $O(N)$ models as a testbed. We compute  the entire momentum 
dependence of the two-point functions in two and 
three dimensions and  obtain excellent results for
both universal and non-universal quantities.

We start by a brief outline of the formalism. In order to simplify the presentation, 
we shall write only the equations corresponding to the case of a scalar field theory 
with quartic coupling,  i.e., restrict the presentation to the case $N=1$ (corresponding to the Ising model).
The strategy of the renormalization group is to build a family of theories indexed by a 
momentum scale parameter $k$,   
such that fluctuations are smoothly taken into account  as $k$ is lowered 
from the  microscopic scale $\Lambda$ down  to 0. 
In practice, this is achieved by adding  to the original  
Euclidean action $S$ a mass-like term of the form   
$\Delta S_k[\varphi]= \frac{1}{2} \int_q\: R_k(q^2)\varphi(q)\varphi(-q)$. 
The cut-off function $R_k(q^2)$ is chosen so that $R_k(q^2)\sim k^2$ for $q\lesssim k$, which 
effectively suppresses the modes $\varphi(q\lesssim  k)$, and so that 
it vanishes for $q\gtrsim k$, leaving 
the modes $\varphi(q\gtrsim k)$ unaffected. 
One then defines a  scale-dependent partition function \be
{\cal Z}_k[J] = \int\! {\cal D}\varphi\; 
e^{-S[\varphi]- \Delta S_k[\varphi] +\int\! J\varphi} \;, 
\label{zk}
\ee
and a scale-dependent 
effective action $\Gamma_k[\phi]$ through a (slightly modified) Legendre transform \cite{Berges00},
\be
\Gamma_k[\phi] +\log {\cal Z}_k[J] = 
\int\! J \phi -\frac{1}{2} \int_q R_k(q^2) \phi_q\phi_{-q}\,, 
\label{legendre}
\ee
 with $\phi=\delta \ln{\cal Z}_k/\delta J$. The variation of the effective action $\Gamma_k[\phi]$ as $k$ varies is governed by  Wetterich's 
equation\cite{Wetterich92}:
\be
\partial_k\Gamma_k = \frac{1}{2} \int_q \partial_k{R}_k(q^2)\, G_k[q,\phi] \;,
\label{rgexact}
\ee
where $G_k[q,\phi]=(\Gamma^{(2)}_k[q,\phi]+R_k(q^2))^{-1}$, 
and $\Gamma^{(2)}_k[q,\phi]$ is the second functional derivative of 
$\Gamma_k[\phi]$ w.r.t. $\phi$. 
The initial conditions of the flow equation (\ref{rgexact}) correspond to 
the microscopic scale $k=\Lambda$ where all  fluctuations are frozen by the 
$\Delta S_k$ term, so that $\Gamma_{k=\Lambda}[\phi]=S[\phi]$. 
The effective action of the original theory is obtained as the solution of  (\ref{rgexact})
for $k\to 0$ where $R_k(q^2)$ vanishes.
Differentiating Eq.~(\ref{rgexact}) $m$ times with respect to $\phi$
yields the flow equation for the vertex function 
$\Gamma^{(m)}_k[q_1,\dots,q_m;\phi]$. Thus for instance,  the flow equation for $\Gamma^{(2)}$ reads:
\begin{equation}
\begin{array}{l}
\partial_k\Gamma^{(2)}_k(p)= 
\int_q \partial_k{R}_k(q^2) G_k^2(q)\left[
\Gamma^{(3)}_k(p,\!-p\!-\!q,q)\times\right. \\ 
\;\;
\left. G_k(p\!+\!q)\Gamma^{(3)}_k(\!-p,p\!+\!q,\!-q)
\!-\! \frac{1}{2}\Gamma^{(4)}_k(p,\!-p,q,\!-q)\right].
\end{array}
\label{rggamma2}
\end{equation}
 (Here we assumed the field $\phi$ to be uniform,
 and omitted the $\phi$ dependence to alleviate the notation.)
Note that the flow equation for $\Gamma^{(m)}_k[q_1,\dots,q_m;\phi]$ involves 
$\Gamma^{(m+1)}_k$ and $\Gamma^{(m+2)}_k$, leading to an infinite hierarchy. 

The flow equation (\ref{rgexact}), and the equivalent flow equations for the vertex functions, are exact, but their solution requires, in general, approximations. It is precisely one of the virtues of this formulation of field theory  to suggest approximation schemes that are not easily derived in other, more conventional approaches.  In particular, one  can develop approximation schemes  for the effective action itself, that is, which apply to  the entire set of correlation functions. The  BMW approximation\cite{BMW} is such a scheme. It relies on two observations. 
First, the presence of the cut-off function $R_k(q^2)$  insures
the smoothness of the $\Gamma^{(m)}_k$'s and limits  the internal momentum $q$ in equations such as Eq.~(\ref{rggamma2})
to $q\lesssim k$. In line with this observation, one neglects the $q$-dependence of the vertex functions in the r.h.s. 
of the flow equations (e.g. in $\Gamma^{(3)}$  and $\Gamma^{(4)}$  in Eq.~(\ref{rggamma2})),  
while keeping the full dependence on the external momenta $p_i$.  
The second observation is that, for uniform fields,
$\Gamma^{(m+1)}_k(p_1,\dots,p_m,0,\phi)=
\partial_\phi \Gamma^{(m)}_k(p_1,\dots,p_m,\phi)$, 
which enables one to close the hierarchy of equations. 

At the leading order of the BMW scheme one keeps the non trivial momentum dependence of the two-point function and implements the approximations above  on Eq.~(\ref{rggamma2}), which
becomes:
\begin{equation}
k\partial_k\Gamma^{(2)}_k(p,\phi)  =  J_3(p,\phi) \left({\partial_\phi\Gamma^{(2)}_k}\right)^2
  \!\! -\! \frac{1}{2} J_2(0,\phi)\, \partial_\phi^2\Gamma^{(2)}_k
\label{BMW}
\end{equation}
with 
\begin{equation}
J_n(p,\phi) \! \equiv \! \int_q k\partial_k{R}_k(q^2) \, G_k^{n-1}(q,\phi) G_k(p\!+\!q,\phi)\;.
\end{equation}
The approximation can be systematically improved: The order $m$  consists in keeping the full momentum 
dependence of $\Gamma^{(2)}_k,\dots, \Gamma^{(m)}_k$ and truncating that of  
$\Gamma^{(m+1)}_k$ and $ \Gamma^{(m+2)}_k$ along the same lines as those leading to
Eq.~(\ref{BMW})  corresponding  to $m=2$.

The zeroth order  approximation is the so-called
local potential approximation (LPA) 
where vertex functions are obtained as derivatives of the effective potential $V_k$ 
(equal, to within a volume factor, to $\Gamma_k$ evaluated for a uniform 
$\phi$), $\Gamma^{(m)}_k(p_1,\cdots,p_m,\phi)\overset{{\scriptscriptstyle{\rm LPA}}}{=} 
V_k^{(m)}(\phi)$, except for 
$\Gamma^{(2)}_k(p,\phi)\overset{{\scriptscriptstyle{\rm LPA}}}{=} p^2 + V^{(2)}_k(\phi)$.  
The LPA has been widely used with  reasonable success \cite{Berges00,delamotte03,canet04b,canet05a}. 
It can be improved through a systematic expansion in gradients of the fields, 
usually referred to as  the derivative expansion (DE) \cite{Berges00,canet03a}. 
However, in contrast to the BMW scheme,  the DE, at any finite order, does {\it not} give access to
correlation functions with non-vanishing external momenta 
(or with external momenta larger than the smallest mass). 

We now turn to the main purpose of the present Letter, which is to show that the 
nonlinear integro-partial-differential 
equation (\ref{BMW}) can be studied as is, without further approximation \cite{NOTE}. 
Note that the earlier  studies of Eq.~(\ref{BMW}) presented in \cite{BMWnum} 
involve additional approximations which are linked to a specific cut-off function, 
and which become too crude  below three dimensions.

In order to treat efficiently  the low (including  zero) momentum sector,
we work with dimensionless and renormalized quantities.
Thus, we measure all momenta  in units of $k$: 
$\tilde{p}=p/k$. We also rescale $\rho\equiv \frac{1}{2}\phi^2$  according to 
$\tilde{\rho}=k^{2-d} Z_k K_d^{-1}\,\rho$ ($K_d=(2\pi)^{-d} S_d/d$, $S_d$ 
being the volume of the unit sphere), and set  $\tilde\Gamma^{(2)}_k(\tilde{p},\tilde{\rho})=
k^{-2}Z_k^{-1}\Gamma^{(2)}_k(p,\rho)$. 
The running anomalous dimension $\eta_k$ is defined by 
$k\,\partial_k Z_k= -\eta_k Z_k$, so that 
at a fixed point $Z_k\sim k^{-\eta}$, $\eta$ being the anomalous dimension 
of the field at the fixed point. The absolute normalization of  $Z_k$ is fixed by 
choosing a point $(\tilde{p}_0,\tilde\rho_0)$ where 
$\partial_{\tilde{p}^2}\tilde{\Gamma}^{(2)}_k\vert_{\tilde{p}_0,\tilde{\rho}_0}=1$.
Then, the flow equation of 
$\tilde\Gamma^{(2)}_k(\tilde{p},\tilde{\rho})$ follows trivially from 
Eq.(\ref{BMW}). 
For  numerical reasons,
we actually solve two equations: one for
$\tilde{Y}_k(\tilde{p},\tilde{\rho}) \equiv \tilde{p}^{-2} 
[\tilde{\Gamma}^{(2)}_k(\tilde{p},\tilde{\rho})-
\tilde{\Gamma}^{(2)}_k(0,\tilde{\rho})]-1$ and one for
the derivative of the dimensionless effective potential
$\tilde{W}_k(\tilde\rho)=Z_k^{-1} k^{-2} \partial_\rho V_k(\rho).$
 Note that 
 $\tilde{\Gamma}^{(2)}_k(0,\tilde{\rho})=
\tilde{W}_k(\tilde\rho) + 2 \tilde\rho\,\tilde{W}_k'(\tilde\rho) $.  
(Here and below, primes denote derivative w.r.t.  $\tilde\rho$.)
These two equations read  (dropping the  $k$ index to simplify the notation):
\begin{eqnarray}
\!\!\partial_t \tilde{Y} &=&  \eta_k (1+\tilde{Y})+\tilde{p}\, \partial_{\tilde{p}} \tilde{Y} -(2-d-\eta_k)\tilde{\rho}\,\tilde{Y}'\nonumber\\ 
&&\!\!\! + 2{\tilde{\rho}}\,\tilde{p}^{-2}\left[(\tilde{p}^2\,\tilde{Y}'\!+\!\tilde\lambda_k)^2 \tilde{J}_3(\tilde{p},\tilde{\rho})
-\tilde\lambda_k^2 \tilde{J}_3(0,\tilde{\rho})\right] \nonumber\\
&&\!\!\! - \tilde{J}_2(0,\tilde{\rho})(\tilde{Y}'/2+\tilde{\rho}\,\tilde{Y}'')\label{eqY} \\
\partial_t \tilde{W} &=&  (\eta_k \!-\! 2) \tilde{W} +(d \!-\! 2 \!+\! \eta_k)\tilde{\rho}\,\tilde{W}'+ \frac{1}{2} \tilde{J}_1'(0,\tilde{\rho}).
\end{eqnarray}
Here $\partial_t=k\partial_k$,
$\eta_k$ is obtained by setting 
$\tilde{Y}_k[\tilde{p}_0,\tilde{\rho}_0]=0$ in Eq.(\ref{eqY}), 
$\tilde{J}_n(\tilde{p},\tilde{\rho})= K_d^{-1}Z_k^{n-1}k^{2n-d-2}{J}_n({p},{\rho})$ 
and $\tilde{\lambda}_k(\tilde\rho)=
3\tilde{W}_k'(\tilde\rho)+2\tilde\rho\,\tilde{W}_k''(\tilde\rho)$.

In practice, we use a fixed, regular, $(\tilde{p},\tilde{\rho})$ grid and restrict the range of the
cut-off function by setting $R_k(\tilde{q}\ge 4)=0$. 
When computing the 
double integrals $\tilde{J}_3(\tilde{p},\tilde{\rho})$, we need to evaluate $\tilde{Y}$
for momenta $\tilde{p}+\tilde{q}$  beyond $\tilde{p}_{\rm max}$, the maximal value on the grid. 
In such cases, we set $\tilde{Y}(\tilde{p})=\tilde{Y}(\tilde{p}_{\rm max})$, an approximation
checked to be excellent for $\tilde{p}_{\rm max}\ge 5$.
To access the full momentum dependence, 
we also calculate $\Gamma^{(2)}_k(p,\tilde{\rho})$ 
 at a set of fixed, freely chosen, external $p$ values. For a given such $p$, $p/k$ is
within the grid at the beginning of the flow. 
This is no longer so when $k<p/\tilde{p}_{\rm max}$; then, we switch to the 
dimensionful version of (\ref{eqY}), and  also set 
$J_3(p,\tilde{\rho})=G(p,\tilde{\rho})J_2(0,\tilde{\rho})$, 
an excellent  approximation when $p>k\,\tilde{p}_{\rm max}$.

We found that the simplest time-stepping (explicit Euler), a
finite-difference evaluation of derivatives on a regular 
$(\tilde{p},\tilde{\rho})$ grid, and the use of Simpson's rule to 
calculate integrals, are sufficient to produce stable and 
fast-converging results. For  all the 
quantities calculated, the convergence to three significant digits  is reached with  
a $(\tilde{p},\tilde{\rho})$ grid of $50\times60$ points; with such a grid, a typical run takes a few minutes on a current personal computer.

 Physical quantities exhibit a small
dependence  on the shape of $R_k(q^2)$ and on 
the point $(\tilde{p}_0,\tilde{\rho}_0)$ where $\eta_k$ is computed.  
Since in the absence of any approximation, they would be strictly independent of the 
cut-off function and of the choice of the renormalization point, 
 a study of this spurious dependence  provides an indication of the quality of the 
present approximation. 
To this end, we use the 
family of cut-off functions $R_k(q^2)= \alpha Z_k\, q^2/(\exp(q^2/k^2)-1)$ 
and vary systematically the parameters $\alpha$, $\tilde{p}_0$, and $\tilde{\rho}_0$. 
In all cases studied,
we find the dependence on $\tilde{p}_0$ and $\tilde\rho_0$ to be much smaller than that
on $\alpha$, so that only the latter needs to  be considered. 
As a function of $\alpha$, 
physical quantities typically exhibit 
a single extremum  $\alpha^*$, located near $\alpha=2$, which moreover, always points towards the best numerical estimates.
Following the principle of minimal sensitivity (PMS) \cite{Stevenson}, 
we regard these extremum values, being locally independent of $\alpha$, 
as our best values. 

\begin{figure}[tp]
\epsfxsize=8.6cm
\epsfbox{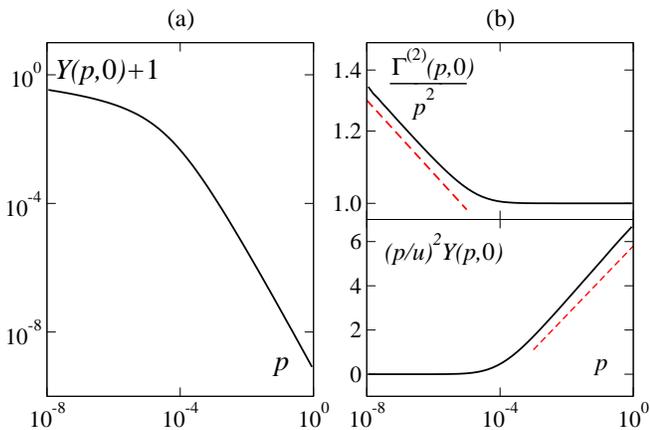}
\caption{(Color online) Typical results at criticality and $k=0$ for $N=2$ in $d=3$ 
($p$ and $u$ are measured in units of $\Lambda$, $u=3\pi^2 10^{-5}$, 
$r_{\rm c}\simeq 6.802854731032857$, $\alpha=2.25$).
(a) dimensionful function $Y(p,0)+1=\Gamma^{(2)}(p,0)/p^2$;
(b) same data as (a). 
Top panel: $\Gamma^{(2)}(p,0)/p^2\sim p^{-\eta}$ with $\eta\simeq 0.041$ for $p\to 0$ (red dashed line).
Bottom panel: expected UV scaling for the self-energy; the dashed line shows the exact two-loop
result for the slope.}
\label{fig}
\end{figure}

We now turn to the discussion of results obtained,  at criticality, first 
in dimension $d=3$ and for various values of $N$.
The initial condition  of the flow is taken to be 
$\Gamma^{(2)}_\Lambda(p,\rho)=  p^2 + r + u \rho$, where the bare coupling $u$ sets a scale, $[u]=[p]$, 
independent of the cut-off $\Lambda$.
Keeping $u$ fixed, the critical value $r=r_{\rm c}$ is found by dichotomy. 
All expected features of $\Gamma^{(2)}_k(p)$ at criticality 
are observed, as illustrated in Fig.~\ref{fig}. Fig.~\ref{fig}a shows the typical shape of $Y(p,0)$.
In the infrared (IR) regime $k\ll p \ll u$, $\Gamma^{(2)}_k(p,0)=p^2(Y(p,0)+1)\sim p^{2-\eta}$,
(Fig.~\ref{fig}b, upper panel). This IR behavior of 
$\Gamma^{(2)}_k(p)$ can be used to extract the value of  $\eta$; 
the value thus obtained is in excellent agreement with that deduced directly 
 from the renormalization condition. The ultraviolet (UV) regime $k,u\ll p \ll \Lambda$ exists 
if $u$ is sufficiently small; this regime  
can be studied perturbatively and one finds that, in leading order, $p^2 Y(p,0)\sim u^2 \log (p/u)$. 
The present approximation reproduces this logarithmic behaviour with, however, 
a prefactor 8\% larger than the  two-loop result
(Fig.~\ref{fig}b, lower panel). Note that the complete two-loop behavior can be 
recovered by a simple improvement of the BMW scheme \cite{Benitez09}.

A quantity particularly sensitive to the  UV-IR crossover region is  the shift, due to interactions,  of the critical temperature of the dilute Bose gas \cite{Baym00}.  In the limit of small coupling, the shift is proportional to $u$. The proportionality coefficient is given (with the normalization used in Ref.~  \cite{Baym00}) by the non-universal quantity
\be
c\,= \left[  -\frac{256}{uN} \zeta[{3}/{2}]^{-\frac{4}{3}} \,
\!\int\! {\rm d}^3 p \left ( \frac{1}{
\Gamma^{(2)}(p)} \!-\! \frac{1}{p^2}\right)\right], 
\ee
in the limit $u\to 0$. Note that the integrand is peaked at values of $p\sim u$, and  is significant for momenta typically  in the range  $[N u/100,10 N  u]$.
Initially introduced for $N=2$, 
corresponding to Bose-Einstein condensation, $c$ is often used as a 
sensitive benchmark of various approximations, as it tests the 2-point function over a wide range of momenta. It has been 
computed,  for several values of $N$, on the lattice and with high-order 
(six-loop) perturbation theory (see Table \ref{table}).

Table \ref{table} contains 
our results for $c$ and the critical exponents  $\eta$, $\nu$ and $\omega$,
together with some of the best estimates available in the literature.
Our numbers are all given for the PMS values $\alpha^*$ of the cut-off parameter, 
and the digits quoted  remain stable  when 
$\alpha$ varies in the  range $[\alpha^*-\frac{1}{2},\alpha^*+\frac{1}{2}] $.
The quality of these numbers is obvious: 
For all $N$ values where six-loop resummed calculations exist, our results for 
$c$  are within the error bars (and comparable to those obtained from 
an approximation specifically designed for this quantity \cite{BMW-BE}); the  
results for $\nu$ agree with previous 
estimates to within less than a percent, for all $N$; as for the values of  $\eta$ and $\omega$, they 
are typically at the same distance from the 
Monte-Carlo and temperature series estimates as the results from resummed perturbative calculations. 
For $N=100$, we find $c=2.36$, $\eta=0.0023$, and $\nu=0.990$, which compare well to the exact large $N$ value $c\simeq 2.33$ \cite{Baym00} and to
the  values $\eta=0.0027$ and $\nu=0.989$ obtained in the $1/N$ expansion \cite{Moshe03}.
Our numbers also compare favorably with those obtained at order 
$\partial^2$ in the DE scheme\cite{canet03a}. 

\begin{table*}[tp]
\caption{\label{table} Coefficient $c$ and critical exponents of the $O(N)$ models for $d=3$.}
\begin{ruledtabular}
\begin{tabular}{cllllclllllclllll}
$N$& \multicolumn{4}{c}{BMW} &\ \  & \multicolumn{5}{c}{Resummed perturbative expansions}&\ \ &\multicolumn{5}{c}{Monte-Carlo and high-temperature series} \\
   & \multicolumn{1}{c}{$\eta$} & \multicolumn{1}{c}{$\nu$} & \multicolumn{1}{c}{$\omega$}  & \multicolumn{1}{c}{$c$} &
   & \multicolumn{1}{c}{$\eta$} & \multicolumn{1}{c}{$\nu$} & \multicolumn{1}{c}{$\omega$}  & \multicolumn{1}{c}{$c$} & \multicolumn{1}{c}{Ref.\footnotemark[1]} &
   & \multicolumn{1}{c}{$\eta$} & \multicolumn{1}{c}{$\nu$} & \multicolumn{1}{c}{$\omega$} &  \multicolumn{1}{c}{$c$} & 
\multicolumn{1}{c}{Ref.\footnotemark[1]} \\ \hline
0 & 0.034 & 0.589 & 0.83 & & & 0.0284(25) & 0.5882(11) & 0.812(16) & & \cite{Guida98} &
   &0.030(3) & 0.5872(5) & 0.88 & & \cite{grassberger}\cite{Peli-rev} \\
1  & 0.039 & 0.632 & 0.78 & 1.15 & & 0.0335(25) & 0.6304(13) & 0.799(11) & 1.07(10) & \cite{Guida98}\cite{Kastening03} &
 & 0.0368(2) & 0.6302(1) &0.821(5) &1.09(9) & \cite{Deng03}\cite{Sun02} \\
2 & 0.041 & 0.674 & 0.75  &1.37 & & 0.0354(25) & 0.6703(15) & 0.789(11) & 1.27(10)& \cite{Guida98}\cite{Kastening03} &
   & 0.0381(2)& 0.6717(1) &0.785(20) & 1.32(2) & \cite{Campostrini06}\cite{Arnold01}\\
3  & 0.040 & 0.715 & 0.73 & 1.50 & & 0.0355(25) & 0.7073(35) & 0.782(13) & 1.43(11)&\cite{Guida98}\cite{Kastening03} &
  & 0.0375(5)& 0.7112(5) & 0.773&  &\cite{Campostrini01,Hasenbusch01} \\
4 & 0.038 & 0.754 & 0.72 & 1.63  & & 0.035(4)& 0.741(6)  & 0.774(20) & 1.54(11)& \cite{Guida98}\cite{Kastening03}&
    & 0.0365(10) & 0.749(2) &0.765 & 1.6(1)  &\cite{Hasenbusch01}\cite{Sun02}\\
10&0.022 & 0.889 & 0.80 &&& 0.024 & 0.859 &  &&  \cite{Antonenko98} &&&&&&\\
\end{tabular}
\end{ruledtabular}
\footnotetext[1]{The first reference is for the critical exponents, the second for $c$.}
\end{table*}

The two-dimensional case, for which exact results exist,  
provides an even more stringent test of the BMW scheme. We focus here on the Ising 
model $N=1$ which exhibits  a standard 
critical behavior  in $d=2$, and the corresponding critical exponents. (The coefficient $c$ is  not defined in $d=2$.) The perturbative method that works well in $d=3$ fails here:
for instance,  the fixed-dimension expansion that provides the best results in 
$d=3$ yields, in $d=2$ and at five loops, $\eta=0.145(14)$ \cite{pogorelov07} in  
contradiction with the exact value $\eta=\frac{1}{4}$ \cite{NOTESOKAL}.
We find instead $\eta=0.254$, $\nu=1.00$, and $\omega=1.28$
in excellent agreement with the exact values $\eta=\frac{1}{4}$, $\nu=1$ and the 
conjectured value $\omega=\frac{4}{3}$ \cite{ZINN}.

To summarize, our results show that the single equation (\ref{BMW}) 
(and its generalization to $O(N)$ models) is sufficient to obtain the 
momentum dependence of the two-point function with excellent accuracy, 
in all momentum regimes, for all $N$, and in any dimension. 
 All this is obtained at a modest numerical cost using simple numerical techniques.
The study presented here is only the leading-order of a 
systematic approximation scheme. A study of the higher orders would be necessary in order to quantify the accuracy that has been reached. However,  
the robustness of our results can already be gauged from the weak residual 
dependence on the cut-off function.

We focused  here on critical theories since numerous and accurate results exist for the critical regime, allowing for  detailed and systematic checks,  but it is clear that the method can be also used to
deal with generically simpler situations. For instance, one 
could calculate the structure factor as a function 
of the momentum and the correlation length, which is of experimental  interest.
The effect  of an external magnetic field could also be investigated by taking advantage of
the built-in  field dependence of $\Gamma^{(2)}_k$.
A detailed investigation of the $d=2$, $N>1$ cases, is also at hand. Finally, 
this approach is not limited to 
$O(N)$ theories. It
can also be applied to disordered, nonequilibrium, or quantum systems, 
expanding from existing studies within the DE scheme of, e.g. absorbing phase transitions
\cite{canet04b,canet05a}, or random-field models \cite{tarjus}.

We acknowledge support from the ECOS project \#U05E01 and 
the PEDECIBA program (Uruguay). FB, RMG, and NW thank LPTMC (Paris) for hospitality.


\begin{thebibliography}{50}
\bibitem{NOTE} For an $N$-component field $\vec\phi$,
Eq.~(\ref{rgexact}) involves a trace on the field index and 
there are two kinds of two-point functions, the transverse 
and the longitudinal ones \cite{BMWnum}.
\bibitem{NOTESOKAL} It has been conjectured (see \cite{Peli-rev} and references therein), and this is 
confirmed by $1/N$ calculations, that the presence of non-analytic terms in the flow of the 
$\phi^4$ coupling $u$ could be responsible for the discrepancy between exact and perturbative 
results in $d=2$. According to Sokal, no problem should arise when  all 
couplings, including the irrelevant ones, are retained in the RG flow, as done here. 
This probably explains the quality of our results in $d=2$.
\end{thebibliography}
\end{document}